# Instructional Strategies that Foster Effective Problem-Solving

Chandralekha Singh
clsingh@pitt.edu
Department of Physics and Astronomy, University of Pittsburgh, USA
ORCID#:0000-0002-1234-5458

Alexandru Maries
mariesau@ucmail.uc.edu
Department of Physics, University of Cincinnati, USA
ORCID#:0000-0001-5781-7343

Kenneth Heller
heller@umn.edu
School of Physics and Astronomy, University of Minnesota, USA
ORCID#:0000-0002-4944-0563

Patricia Heller
helle002@maroon.tc.umn.edu
Department of Curriculum and Instruction, University of Minnesota, USA
ORCID#:0000-0002-5102-2440

Helping students become proficient problem solvers is a major goal of many physics courses from introductory to advanced levels. In fact, physics has often been used by cognitive scientists to investigate the differences between the problem-solving strategies of expert and novice problem solvers because it is a domain in which there is reasonably good agreement about what constitutes good problem-solving. Since the laws of physics are encapsulated in compact mathematical form, becoming an expert physics problem solver entails learning to unpack and interpret those physical laws as well as being able to apply them in diverse situations while solving problems. A physics expert must have a well-organized knowledge structure of relevant physics and math concepts and be able to manage cognitive load and do metacognition while solving complex problems. In this chapter, we review foundational research on expertise in physics problem-solving and then discuss research on instructional strategies that promote effective problem-solving as well as challenges in changing the instructional practices of physics instructors and teaching assistants via professional development to promote and support effective problem-solving approaches.

## Introduction

Physics is the quest to understand our natural environment by finding and solving the problems raised by its investigations. This makes problem-solving central to every physics course. Furthermore, the process of solving problems is itself inextricably entangled with all meaningful learning (Shuell, 1990). In this chapter, we give a brief, and necessarily incomplete, survey of the research basis of instructional strategies that improve student problem-solving and examples of some of these strategies. The foundation of problem-solving instruction comes from work in the fields of philosophy, education, psychology, neuroscience, and computer science whose interconnections comprise the interdisciplinary field of cognitive science. In our references, we have chosen work that is accessible to practitioners of Physics Education Research (PER) and provide a chain of citations to the origins the research foundations of problem-solving instruction. The challenge for PER investigators is to determine how instruction that fosters effective problem-solving can be routinely implemented by real instructors in real physics classes with real students.

In the following sections of this chapter, Problem Solving: A Brief History Leading to PER; What is a problem; What is a problem?;Factors that Impact Problem Solving; Building Practical Instruction; Incorporating Problem-solving into Instructional Design; Consideration of Motivational Factors for Fostering Student Attitudes about Problem Solving; Professional development of TAs and instructors: Beliefs about Problem Solving; Relate Instructional Framework; Conclusion; and References.

Of course, instruction does not occur in a vacuum but is constrained by institutional considerations and the need to change both instructor practices and student expectations. We point out some of the challenges posed by common instructional practices that undermine the learning of effective problem-solving and the need for instructor professional development to address them.

## Problem Solving: A Brief History Leading to PER

Our starting point is the cognitive science definition of problem-solving as "the process of moving toward a goal when the path to that goal is uncertain" (Martinez, 1998). This process entails refining the formulation and manipulation of conceptual knowledge starting with clarifying a testable goal and finishing with determining the validity of the path constructed. Along the way, the solver assembles, modifies, and constructs ideas that might prove useful and ignores those that might not. This is a recursive effort to build and execute a method to achieve the goal. The process of solving problems is an act of individual creation that goes beyond remembering facts and applying recipes. It is a personal process that requires activating the brain's existing neural pathways, building new neural connections, and remodeling their linkages. This is the very definition of learning. In the lifelong evolution of a person's problem-solving their problem-solving ability changes as their brain develops and their brain develops as problem-solving occurs (Rees et al, 2016). As with any human accomplishment, problem-solving is situated in the rich experience of the solver and develops over decades with experience and instruction (Dreyfus, 2004).

The written history of understanding the human action of problem-solving goes back at least as far as Aristotle (Quarantotto, 2020). Skipping over important developments in the next couple of thousand years, to the early 20th century, we reach the influential work of Dewey (1910), Piaget (1967), Vygotsky (1978), and Polya (1945). Their work illustrates the utility of qualitative research that uses case studies, artifact driven interviews, and anthropological observation to drive personal introspection. By the late 20th century, advances in understanding brain functioning, exemplified by the work of Hebb (1949), coincided with the development of artificial intelligence, pioneered by Turing (1950) to stimulate investigations of the mechanisms of problem-solving in the new field of cognitive science. From early days of cognitive science, some of seminal advances in problem-solving used physics as a context (e.g. Simon, 1974, Simon and Simon, 1978, Larkin et al., 1980, Chi et al., 1981).

To uncover what comprised problem-solving, researchers began comparing the approaches of strong problem solvers, called experts, with weak problem-solvers, called novices (Newell et al., 1958, Chase & Simon, 1973, de Jong and Fergusson-Hessler, 1973, Simon, 1974, Larkin & Simon, 1987, Chi, et al., 1981, Heller & Reif, 1982, 1984, Eylon & Reif, 1984, Sweller, 1988, Simon & Simon, 1978, Bassok & Holyoak, 1989). It was found that the difference between experts and novices was much more than a difference in their amount of knowledge. Experts solved problems using very different cognitive processes than novices and even attended to different information in the problem situation (Bransford, 2000).

As with any early research, the pioneering studies had procedural difficulties. For example, having both novices and experts solve the same question seems to give a direct comparison. However, a physics question that is a tractable problem for a novice, is not a problem for an expert. For the expert, the solution path may have choices, but it is not uncertain. One could give the novice and expert a question outside of the domain of either, but expertise is domain specific. Then there is the difficulty of determining who is an expert in the domain. For example, experienced physics professors are usually considered experts in solving problems in the physics domain. However, is a physics graduate student, who is almost certainly better at solving a physics problem than a freshman, an expert? Is a person either a novice or an expert problem solver?

It is now known that there is a range of problem-solving performance that goes from beginner to expert and beyond (Dreyfus, 2004). The range has been characterized into intermediate stages of novice, advanced beginner, competent, proficient, and expert. This classification is not unique and is neither discrete nor linear nor continuous. However, it gives a language for discussing instruction. Moving from novice to expert in any endeavor takes at least a decade of deliberate practice (Ericsson et al, 1993), a timescale well beyond the reach of single class or even an educational institution's curriculum. That means that instruction can help a student progress toward expertise, but they will probably not achieve expertise while still in our educational system. Assuming optimal instruction within the educational system, we would expect most experienced high school physics teachers or physics professors to be either expert or proficient physics problem solvers, most graduate student teaching assistants to be competent or proficient, and most students who are beginning to learn physics to be novice or advanced beginner.

## What is a problem?

The definition of problem-solving as "the process of moving toward a goal when the path to that goal is uncertain" (Martinez, 1998) assumes there is a goal. However, this definition encompasses a great deal of complexity (Anderson, 2000). What makes the path toward a goal uncertain is usually that the goal is initially perceived as ill-defined. Clarifying or revising it is often the most important part of problem-solving, a process that continues until a solution is reached and justified. This goal clarification entails recognizing the essential features of the situation and separating them from the inessential, or surface, features (Chi et al, 1981).  Once the solver decides on the essential features of the situation, they can decide how these features are constrained by the situation of the problem.

Whether or not a question is a problem depends not upon the complexity of its solution, but on the perspective of the problem solver because path uncertainty is in the eye of the beholder.  This results in a classification of goal directed questions into two categories: problems and exercises.  An exercise has a goal that is clear to the solver attached to possible choices of solution path.  The exercise may be difficult to resolve requiring many steps to reach a solution and even entail leaning new techniques.  Exercise solving is an important tool for learning and automating skills needed for problem-solving.  However, exercise solving by itself does not provide the experience needed to solve problems.

## Factors that Impact Problem Solving

Helping students develop toward expertise in physics, requires appreciating three interrelated factors that impact problem-solving: metacognition, knowledge organization, and information processing.  Figure 1 is a representation of some of the foundational research that investigated how these factors influence instruction. These three factors are essential building blocks of problem solving and all impact a student's brain activity, called cognitive load, when solving a problem. We recognize that our organization of the problem-solving research is only one of many. Indeed, studies that we cite often cut across the organizational boundaries.

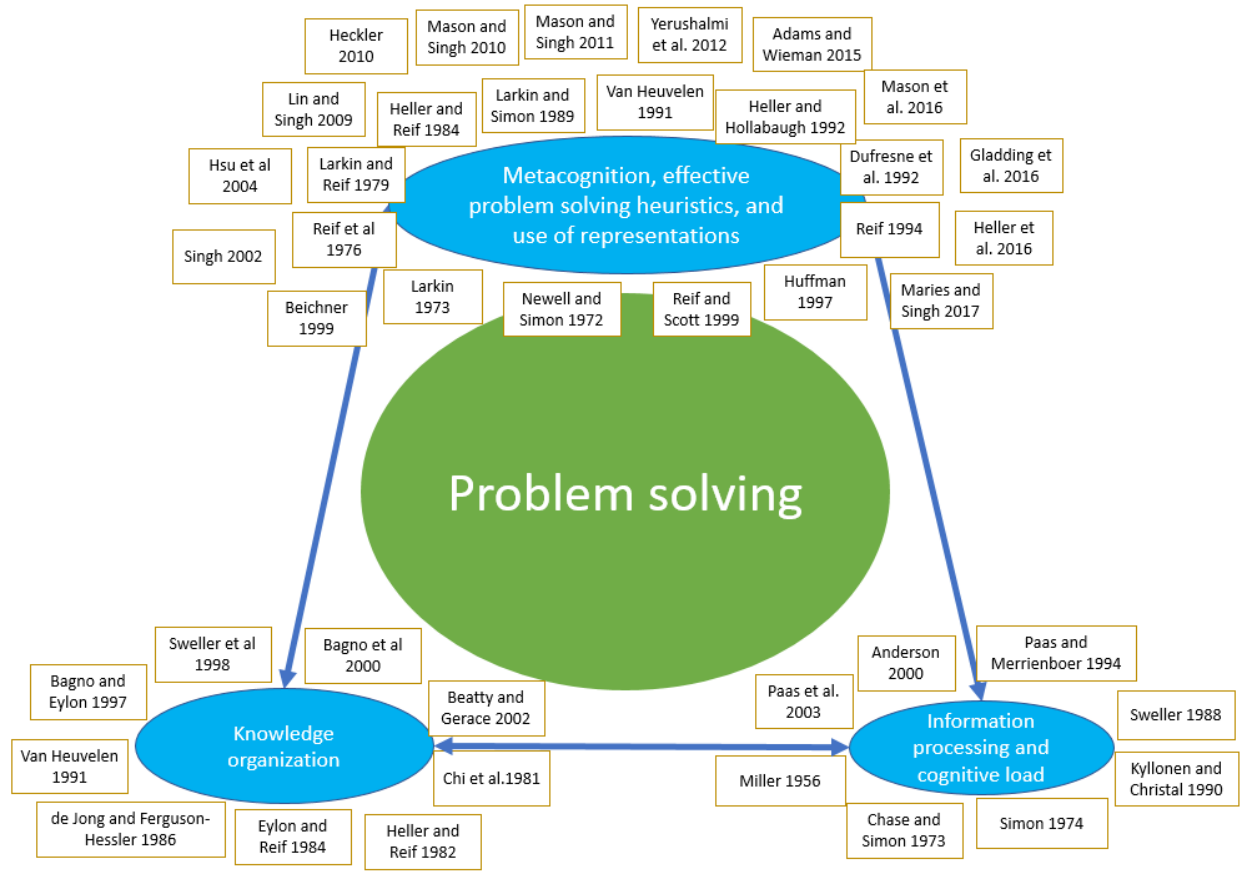

**Figure 1.** Research studies related to the three pivotal issues in physics problem-solving instruction: 1) Information processing and cognitive load, 2) Knowledge organization and 3) metacognition.

## The Role of Metacognition

Because the process of problem-solving is a web of decision-making, a proficient problem-solver needs to be in conscious control of what decisions are needed and how they are made. This process of controlling and monitoring mental decisions is called metacognition (Flavell, 1976, Schoenfeld, 1987, Jacobs & Paris, 1987, Schraw, 1998, Morphew et al., 2020). In fact, problem-solving itself can be called a metacognitive skill (Martinez, 2006). Broadly speaking, the three main metacognitive skills used in problem-solving are planning, monitoring, and evaluating. Planning involves selecting a goal and appropriate strategies to attain that goal before embarking on solving a problem. Monitoring is the awareness of progress toward the goal of a problem solution. Evaluating is constantly appraising the value of the solution. The monitoring and evaluating skills allow the solver to modify the plan or even the goal as the solution progresses. Novice problem solvers do not recognize the need for metacognition when solving physics problems which limits their ability to do so. To progress to the competent level, students must become more metacognitive in their problem-solving process. Some do so naturally, but most need instruction that makes the metacognitive processes explicit. As students become better problem solvers, their use of metacognition has a cost. It adds a significant cognitive load that can limit their proficiency and add to their frustration (Sweller, 2019). Instructional techniques need to be employed that give students practice using metacognition without significantly increasing their cognitive load.

## Information Processing and the Constraint of Cognitive Load

Because problem-solving and metacognition are entangled, cognitive load management becomes an important issue for instruction. Early research in cognitive science hypothesized that there are two broad components of human cognitive architecture (Anderson, 1995): long term memory (LTM) and working memory (WM). Working memory is where information processing occurs and has a finite storage of $7\pm2$ "slots" (Miller, 1956). Each of these slots can contain a single piece of information or a link to an entire concept or process. On the other hand, LTM does not appear to be limited in the amount of information it can store. During the problem-solving process, working memory receives inputs from sensory buffers (e.g., eyes, ears, hands) and LTM (Sweller, 2011). Since the number of WM slots that can be processed at any given time is small, there is an advantage to organizing information into larger "chunks" that each occupy only a single slot (Kyllonen & Chirstal, 1990, Kail & Salthouse, 1994). For example, over time students can develop a large information-processing chunk about how to solve simultaneous equations that only takes up one slot of WM. The mechanism for WM is thought to reside in the neural structure of the brain's prefrontal cortex (Funahashi, 2017)

One finding of Sweller's work (Sweller, 1988) is that, without guidance and support, many students use a disorganized means-end analysis to solve problems. Broadly speaking, this approach entails continuously considering the goal and the current state of the solution relative to that goal. This for novices problem-solving is the process of reducing the distance between the goal and the current state. When well-organized, means-end analysis, or working backwards, is an important expert problem-solving tool. For example, engineers (Dorst K, 2015b) and curriculum designers (Wiggins & McTighe, 2005) call this process backward design. This approach is often contrasted with another expert approach, working forward, that starts with a qualitative analysis of the situation and, by successive translations, converts it into more specific representations that lead to a solution. Working backwards usually entails a larger cognitive load than working forward especially for students at or below the competent level. Experts combine working backwards and forwards to solve problems. However, a novice does not have the hierarchal knowledge structure to work forward.

Working memory activity also appears to play a mediating role in explaining the negative impact of stereotype threat, the threat that women and other underrepresented students often feel while solving physics problems due the anxiety of conforming to societal stereotypes, on problem-solving performance (Beilock et al., 2006, 2007). The problem-solving skills that rely heavily on working memory are heavily impacted by that anxiety due to its processing in working memory (Maloney et al., 2014). The greater the anxiety, the less space is available in the WM for the needed problem-solving information processing.

## The Role of Knowledge Organization

The research on knowledge organization was conducted quite early by cognitive scientists. These early studies used think aloud protocols to determine how knowledge is organized and retrieved by an expert, such as a physics professor, and a novice problem solver (Chi et al., 1981, Heller & Reif, 1982, 1984) solving the same problem. These think-aloud interviews were sometimes explicitly used to recreate a representation of experts' and novices' knowledge structures (for example, see Figure 2). Experts organize their knowledge around the fundamental principles of physics at the top, followed by more ancillary concepts. This hierarchical knowledge organization around the fundamental principles and concepts of physics is efficient at producing the correct approach for solving a problem. Novices have a more random knowledge structure, where fundamental principles are mixed with the surface features of the problem. During novice problem-solving, often it is not that the knowledge is necessarily missing in the student's knowledge structure, but rather how accessible the knowledge is to searches, based on the cues in the problem statement about how to solve the problem.

Another early study of knowledge organization (de Jong & Fergusson-Hessler, 1986) found that students who perform well in physics classes categorize problems according to physics principles (e.g., this is a Newton's Second Law problem), while those that do not do as well categorize the same problems according to the problem situation, so-called surface features (e.g., this is an inclined plane problem). This categorization approach, based on the way knowledge is organized, has also been replicated in mathematics (Schoenfeld & Herman, 1982). More recent research also used problem categorization with large numbers of introductory students and found that those students categorized problems along a continuum (Singh, 2009a, Mason & Singh, 2011). Some categorized the problem very similarly to experts, others were much closer to novices, with the rest on a continuum in between. The study has also been replicated in the context of quantum mechanics (Lin & Singh, 2010). This suggests that well-constructed categorization tasks might be used to help students organize their knowledge around fundamental principles (Docktor et al., 2012, Mason & Singh, 2016b), as well as for assessment (Hardiman et al., 1989, Singh, 2009b, Mason & Singh, 2011).

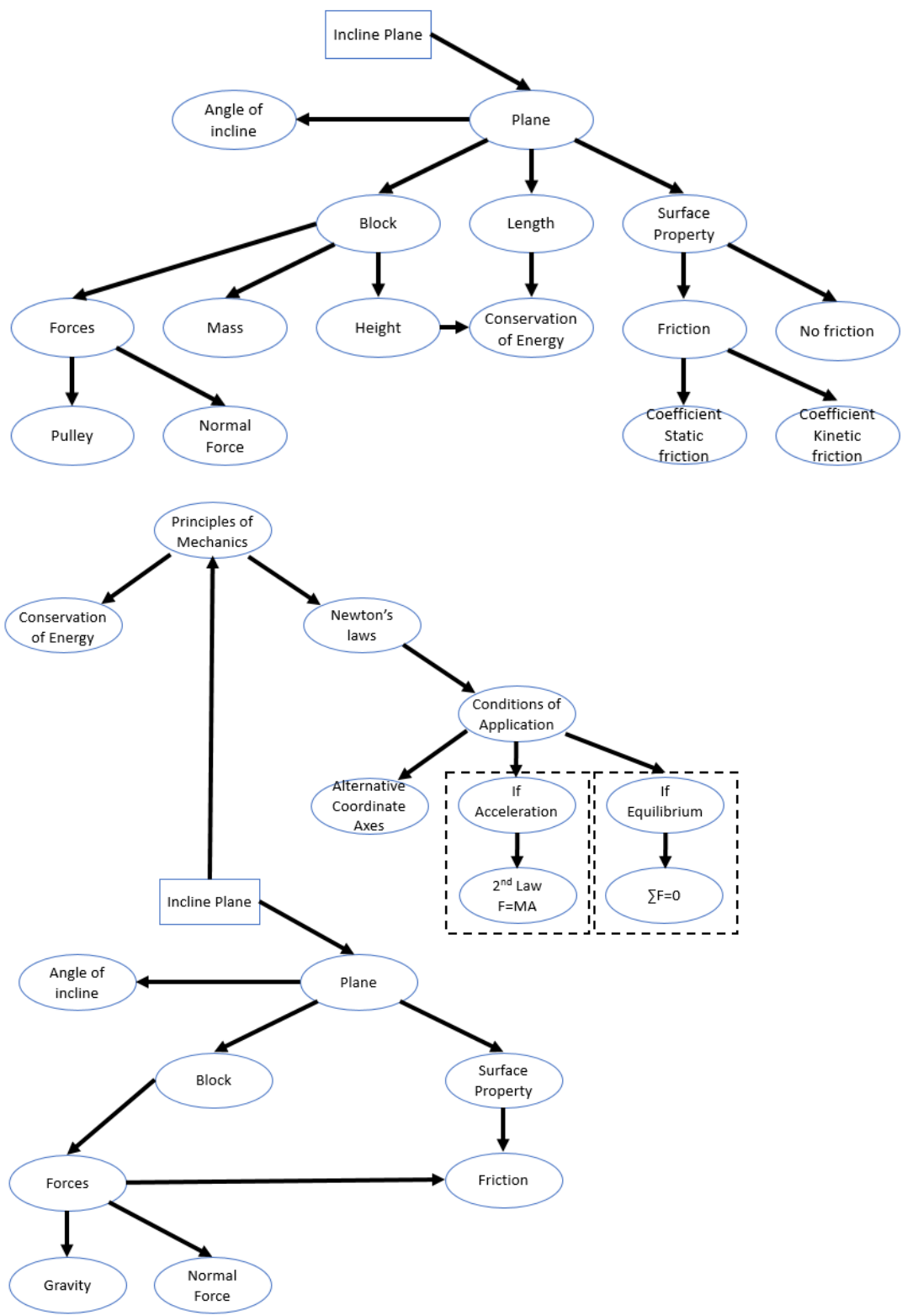

**Figure 2.** Knowledge organization of a novice (top) and an expert (bottom) as inferred from novice/expert think-aloud interviews while solving the same problem from Chi et al., 1981.

Eylon and Reif's (1984) research also revealed the importance hierarchical knowledge organization in improving students' problem solving. In the study, students learned the same content with knowledge presented either hierarchically or sequentially. They found that students performed significantly better in subsequent tasks when taught the content with knowledge presented hierarchically. The researchers concluded that "a hierarchical organization should facilitate performance on complex tasks involving appreciable information retrieval." Others (e.g., de Jong & Fergusson-Hessler, 1986) have similarly argued that having the knowledge is not enough, it must also be organized in a useful manner in order for it to be retrieved appropriately during problem-solving.

Chunking problem solving processes that are used in large categories of problems is called building heuristics. Examples of heuristics, in order of increasing extent, are drawing a free-body diagram, representations such as vectors, using Newton's second law, or classical mechanics. Larger classifications (e.g., classical mechanics) include heuristics embedded within heuristics. As students build increasingly extensive concepts and heuristics, solving problems requires less extensive metacognition and lower cognitive load. In other words, developing expertise in physics requires using problem-solving as a learning opportunity and becoming proficient in using a handful of laws of physics to solve a wide range of problems.

## Building Practical Instruction

The goal of PER is to build a more effective way for all students to learn physics, in this case problem-solving. PER research, directly or indirectly, contributes to achieving that goal. The test of that research is to build measurably successful instructional structures that encompass the diversity of its students and instructors. To describe the progress in building those structures, we outline the current overarching framework of instruction, cognitive apprenticeship. Of course, practical instruction must address the needs of real students and real instructors operating in real institutions. For that reason, we will mention two difficulties that must be addressed by any problem-solving implementation. In the next section (Incorporating Problem-solving into Instructional Design) we give examples successful problem-solving instruction tested with significant numbers of students in an ecologically valid situation. We emphasize that there is not enough space to be inclusive so many important instructional methods have been left out.

## Cognitive Apprenticeship: An Instructional Framework

At this time, cognitive apprenticeship (Collins et al., 1989) is the best general model of instruction. As an empirical model it integrates the features of traditional apprenticeship with the research results of cognitive science to produce a framework for effective instruction. The overarching principle of cognitive apprenticeship is that learning must be situated in a context that is realistic to the student and authentic to the subject matter. Cognitive apprenticeship actions are guided by the instructor's knowledge of their students' current abilities and the extent of their zone of proximal development (Vygotsky, 1978). Instruction using this framework is comprised of four actions: modeling; coaching; scaffolding and fading. For problem-solving, modeling is the action of demonstrating how a student might solve problems that are used for both practice and assessment. Modeling is accomplished by the instructor working an example in front of the class, providing a video, engaging in a Socratic dialog with students in a class, or distributing example solutions. To be effective, the modeling action needs to demonstrate how students could solve the problem, not how the instructor might solve it. In all cases, modeling must make visible to the students the usually invisible mental processes of metacognition and information processing.

Coaching is the personal feedback to a student while they are engaged in solving similar problems modeled in lectures. It builds on a student's actions and cognitive processes as opposed to showing them "the right way to do it". When being coached, a student must receive prompt feedback that could come from peers, an instructor's assistant, or a well-designed computer program. Scaffolding is the temporary educational structure to help students build an intellectual bridge from their current knowledge state to the next achievable level. Scaffolding supplies the "training wheels" to allow students to practice skills that they are not yet ready to initiate. It bridges the gaps in their knowledge and confidence to allow them to practice the next level of problem-solving desired by the instructor and within the student's zone of proximal development. Fading is the gradual reduction of scaffolding, coaching, and modeling with respect to the specific knowledge being constructed (e.g. Heller & Heller 2010). Fading occurs when the trappings of instruction are gradually reduced allowing the student to attempt the desired level of performance on their own.

The cognitive apprenticeship actions are not sequential and not always needed by every student in every instance. For example, after experiencing a model of using a hierarchal knowledge organization, a student might want to try it on their own. If they are unsuccessful, they can take advantage of more coaching or observe more modeling. Within the cognitive apprenticeship framework, problem-solving helps students develop a robust physics knowledge structure that they can use to relate one situation to another (Lin & Singh, 2013a, 2013b, Whitcomb et al., 2021). For example, with appropriate scaffolding, students can learn, over time, to recognize that all classical mechanics problems are essentially the same even though they may have very different surface features and difficulty. The cognitive apprenticeship actions are not sequential and not always needed by every student in every instance. For example, after experiencing a model of using a hierarchal knowledge organization, a student might want to try it on their own. If they are unsuccessful, they can take advantage of more coaching or observe more modeling.

## Physics Concepts and the Mathematics to Express Them

An important aspect of physics is the proficiency with which one makes appropriate connections between physics concepts and the mathematics to express them (Reif, 1995). Unfortunately, mathematical calculations can be tedious and prone to mistakes which makes the small part of physics problem-solving that is mathematics loom large in the minds of students. Allowing students to use modern technology to do algorithmic calculations relives some of the stress caused by calculations while maintaining a focus on the importance of mathematics. For example, almost any complex mathematical process, such as solving a set of simultaneous equations, can be reduced to a single line of software. This calculation trap in instruction has a long been recognized within physics (e.g. Oerlein 1937, Crane 1966, Crane 1969) but its depth was generally not recognized until studied within the framework of metacognition, information processing, and knowledge organization. Most proposed instructional solutions rest on the design of scaffolding that pays attention to problem construction, the guidance of student solution construction, peer coaching (e.g. Heller et al, 1992a), Heller & Hollabaugh, 1992b), computer coaching (e.g. Ryan et al, 2016), grading (e.g. Henderson 2004), and the teaching assistants that provide coaching (Lawrenz et al 1992). However, it is possible that a long-term solution is reducing student calculation while maintaining the mathematical nature of physics by a more extensive student use of computational tools (Caballero et al 2021).

## The Different Strategies of Instructors and Students

The qualitative disparity between instructor and student problem-solving causes difficulties for physics instructors that can lead to what is referred to as the "expert blind spot" (Nathan et al., 2001). This disparity makes it difficult for an instructor to evaluate a problem's difficulty from students' perspective and to scaffold student learning. Because any tractable assignment for the student is not a problem for the instructor, the process by which the instructor constructs an answer usually does not represent the way either they or their students solve problems. This is why instruction designed to lead a student through a process the instructor might use to get an answer often does not give the student the practice needed to attain the next level of proficiency in problem-solving (Singh, 2002, Yerushalmi, et al. (2007)). Even if the instructor addresses the question as if it were a problem to them, many of their own problem-solving processes have become automatic and subconscious (Schraw, 1998, Schoenfeld, 1987, Jacob & Paris, 1987). The instructor is not aware that they use certain problem-solving processes and so cannot make them explicit for their students (Reif, 1995). The result is that the students, do not know that those problem-solving decisions exist which gives rise to the common student experience of "not knowing where to start". To remediate the learning gaps caused by an instructor's expert blind spot, instructors can ask students to think aloud while solving problems and perform a cognitive task analysis (Chipman et al., 2000) of student's problem-solving process. This analysis can help an instructor recognize their own blind spots and provide scaffolding to help the student practice the problem-solving processes necessary to increase their proficiency (Reif, 1995, Dufresne et al., 1992, Hsu & Heller, 2005). This is an example of how research on problem-solving can provide instructors an insight into the frustration students feel when solving physics problems. It can also suggest how an instructor can help students bridge that frustration by identifying missing problem-solving processes.

## Incorporating Problem-solving into Instructional Design

The following instructional strategies have been successfully used to improve students' problem-solving abilities within a physics course. Intentionally or not, they use most of the features of the cognitive apprenticeship model. These strategies are designed to reduce the cognitive load required to solve a problem, to help students develop a hierarchical knowledge organization, and/or develop their metacognitive skills. Many test types of scaffolding for use during modeling and coaching.

## Attending to Students' Conceptual Difficulties

Not surprisingly, most researchers who investigate problem-solving emphasize the importance of the initial qualitative (i.e., conceptual) analysis of the problem (e.g., Heller & Reif, 1984, Reif, 1995, Heller & Hollabaugh 1992a, Beichner 1999, Huffman 1997, Koedinger & Nathan, 2009, Docktor et al 2015, Ryan et al., 2016). However, students have strong ideas about physics that are incorrect (McDermott, 1991). These alternative conceptions or misconceptions interfere with students' ability to successfully solving problems.

Peer Instruction (Mazur, 1997) restructures the introductory physics course to focus on qualitative reasoning by building lectures around conceptual questions that address student misconceptions. Students answer them electronically. After seeing how others answered, the students discuss their result and resubmit their answer. This can then lead to an explanation by the lecturer. The class also emphasized such reasoning on examinations. Student performance on conceptual problems improved significantly. Additionally, if quantitative problems contributed to the course grade, then emphasizing conceptual understanding in lectures also slightly improved the performance on quantitative problems. Similar results have been obtained by replacing traditional recitations, which emphasize quantitative problem-solving, with conceptual tutorials (McDermott, 2001) for traditionally taught classes that heavily focus on quantitative problem-solving.

Another example of attending to students' conceptual difficulties in instructional design is an approach that "integrates problem-solving, conceptual understanding, and the construction of a knowledge structure" (Bagno & Eylon, 1997, Bagno et al. 2000). The goal is to help students organize their knowledge around key concepts and use it to solve problems to develop a hierarchical knowledge organization. One of the key characteristics of the approach is that students are led to construct their knowledge structure through problem-solving and generate concept maps, such as shown in Figure 1, that link the central concepts of the topics. Students solve problems, reflect on the central concepts that relate the problems and represent them in different forms. They then develop and elaborate the concepts to overcome their conceptual difficulties and apply this knowledge to solving a novel problem. To consolidate this knowledge, they link the new concept to a map of their prior knowledge. This implementation found that students benefit from making explicit links (via concept maps) between the concepts of mechanics and electromagnetism. This is similar to the ALPS-kit approach that includes explicit activities where students organize the concepts learned and link them to those from previous activities (Van Heuvelen, 1991a).

## Using Analogical Reasoning

Analogical reasoning can be a useful approach to help students learn physics problem-solving especially when they have strong alternative conceptions (Holyoak, 1985, Novick, 1988, Clement, 1998, Lin & Singh, 2011a, 2011b, Singh, 2008a, 2008b). Comparing two problems that have very similar solutions despite differences in surface features helps students recognize their underlying structure (Lin & Singh, 2011a). When students solve isomorphic problems, it helps reinforce general effective problem-solving strategies (Lin & Singh, 2013a, 2013b, Docktor et al., 2012, Mason & Singh, 2016b). For example, solving a quantitative problem before an isomorphic conceptual problem helps students recognize that most of problem-solving is conceptual and that conceptual problems require the same type of careful reasoning as quantitative ones (Singh, 2008c).

The next section gives examples of scaffolding that allows students to practice important elements necessary to become a better problem solver.  However, this practice will only occur if the grading of students discourages terse or random attempts at problem solutions and rewards explicit evidence of using physics representations, hierarchical knowledge structures, and problem-solving outlines.  Research into instructors' beliefs about problem solving indicates that most physics instructors prefer short, unexplained mostly mathematical solutions and do not reward a well-organized and explained problem solving process (Henderson et al, 2004).

## Developing Useful Representations

In physics, a useful representation depends on the problem context. For example, for introductory mechanics problems involving Newton's laws, Reif suggested that the representation should separate the system of interest from its environment, represent its motion and identify the interactions on it starting with long-range forces that do not require contact identified first and then contact forces identified at their contact points (Reif, 1995).  The system description in turn is used to build equations that represent the physics of the problem.  These equations are then used to solve for the problem goal.

In the Overview: Case Study Physics (OCS) approach (Van Heuvelen, 1991a), the student first practices representing physics concepts using diagrams and graphs.  Only after the student gains sufficient experience working with those representations, do they explore the concepts mathematically and use them in problem-solving (Van Heuvelen 1991b) . This approach allows the student to combine the processes of building representations into heuristics that reduce their cognitive load when later used in problem-solving (Larkin and Simon, 1987).  The representation artifact itself externalizes and thus extending working memory (Zhang & Norman, 1994).  More recently others have used this approach with interactive computer and video-based tutorials. These tutorials improve student problem-solving when used with individual human coaching to motivate the students to complete them.  However, when studying or working homework, students tended not to use them for deliberate practice (Marshman et al., 2020a).  A related implementation (Koenig et al., 2022) combined the process of interactive tutorials with modeling by a video featuring a human that appears as needed.  Although still in a test stage, the on-demand interleaving of electronic coaching and modeling, which is more consistent with cognitive apprenticeship, appears to be more successful at motivating students to persist.

## Developing a Hierarchical Knowledge Structure

The Hierarchical Analysis Tool (HAT) is an example (Dufresne et al., 1992) is a computer tutor that requires students to follow a general hierarchical approach to solving problems.  To begin a problem solution, students are given a series of well-defined questions from which they choose the broad principles that apply to the problem. Their answer to subsequent questions guides them to choose the subprinciples that lead to the solution. As the student makes their choices, the HAT software gives the equations consistent with each choice. It is up to the student to solve the equations produced by this concept analysis. The researchers found significant performance differences in problem-solving between students who used the HAT tool and students who solved problems traditionally, and they conclude that this type of qualitative analysis results in a significant improvement in problem-solving behavior.

The Personal Assistant for Learning (PAL) software (Reif & Scott, 1999) is an example of a computer tutor that guides the student through the entire problem-solving process for Newtonian force problems. It combines the actions of a tutor that guides the student to build physics representations of the problem with one that enforces the use of a hierarchal knowledge structure that results in the student building the equations needed to solve the problem. This is combined with a reciprocal teaching process (Palincsar & Brown, 1983) where the student becomes the tutor for the computer. For this limited implementation, the students using the PAL performed as well solving introductory mechanics problems as those tutored by a human and better than those taught in a regular class.

## Scaffolding the Problem Solution

Providing students with scaffolding (Wood et al, 1976) is a common method of reducing a students' cognitive load caused by the intense information processing and metacognition during problem solving, as a means of navigating Vygotsky's (1978) zone of proximal development. One common scaffolding is providing students with a general problem-solving framework, often introduced as a worksheet (Chi & Glaser, 1985, Hardiman et al., 1989, Heller et al 1992a, Reif, 1995, Dufresne et al, 1997, van Someren et al., 1998, Van Heuvelen & Zou, 2001, Rosengrant et al., 2005, Warnakulasooriya et al., 2007, Singh, 2008c, Nguyen et al., 2010, DeVore et al., 2017, Maries & Singh, 2018a, 2018b). This framework guides students, to approach a problem in a systematic way. There are four steps for short problem statements that contain physics cues such as "find the final momentum": (1) describe the problem in physics terms, including useful representations, and a list of physics knowns, unknowns and constraints; (2) a detailed plan to solve for the goal, including selecting pertinent principles or concepts that might lead to a solution and writing down the logical or mathematical steps that leads to a solution; (3) executing the plan; and (4) checking that the answer makes sense. On the other hand, more realistic problems contain no words that cue the physics concepts and principles needed in its solution, and its quantitative goal may not be well-defined. These problems require a beginning step to bring the problem into focus (Heller et al, 1992a, Heller & Heller (2010). This step involves clarifying the problem by drawing a sketch of the situation including possibly important physics quantities and determining a quantity that once calculated solves the problem. Although laid out in a sequential manner, at each stage in the framework the student is guided to actively monitor the value of their actions in progressing toward their goal and to modify previous actions if progress is stalled.

Once the necessary cognitive apprenticeship action of fading occurs and the problem-solving process is not scaffolded, students should be encouraged to write an outline of their solution process before embarking on it. Rewarding students when grading will encourage outlining even before the scaffolding is removed. This extends their working memory to include the sheet of paper and allows them to practice chunking by categorizing the parts of problem-solving and keeping their cognitive load manageable. (Heller et al, 1992a). Outlining is especially useful in any extended calculation for students who are not use to chains of mathematics or have math anxiety (Foley et al, 2017). This process is similar to writing pseudocode, which is recommended for building software (Hatziapostolou & Paraskakis (2008) and has always been used as a tool for writing.

## Reflection After a Problem is Solved

The metacognitive process of reflection after a task is the mechanism by which people repair, organize, and extend their knowledge structure. Unfortunately, in most instructional settings students have no time or incentive to do so. One method of introducing students to this process is to provide them with meaningful incentives to diagnose their problem-solving mistakes after taking a test or completing an assignment. In the studies of students after taking a test, the best intervention for students to learn from their mistakes was providing them a meaningful reward but only minimal scaffolding. (Yerushalmi et al., 2012a, 2012b). In this experiment, students could correct their mistakes to regain 50% of any points lost on a problem. However, the students that were given a detailed solution to the problem did less well than student who could only use their own resources, textbook and notes). Students provided with too much external support, seemed to perform the reflection task superficially without engaging with the underlying physics concepts. In another study (Brown et al., 2016) showed that advanced students do not necessarily learn from their mistakes even if the correct solution is provided to them.

Another approach to encourage students to reflect on their problem solving after they have solved it is to use reflection with peers (Mason & Singh, 2010a, 2016a). In this application, the incentive is a competition. Students first work individually to solve a problem, then work in groups of three to decide which solution is the "best" one. The chosen solution competes with solutions chosen by other groups. Then in a whole-class discussion a winner is chosen whose solution best exemplifies effective problem-solving strategies. This type of activity results in students using a greater range of effective problem-solving approaches than in more traditional classes.

## Cooperative group problem solving

Cooperative group problem solving is a complete introductory physics class structure that encompasses all the features of cognitive apprenticeship (Heller and Heller 2010). The structure of the course was based on existing research and both quantitative and qualitative data from the students (Heller et al, 1992a, Heller and Hollabaugh, 1992b). The course developed incrementally from its initial goal of helping students develop a generalizable form of problem-solving that emphasized the conceptual structure of physics and its mathematical nature.

Modeling is accomplished in lectures that introduce physics concepts as solutions to problems. Those concepts are combined with previous concepts by the lecturer to demonstrate the solution to problems that are realistic and use natural language to avoid physics cueing, called Context-rich problems. The lecturer demonstrates problem-solving by articulating both the external problem-solving process and the internal decision-making process. Informal peer Coaching occurs in the lectures where students work in groups of neighbors to occasionally determine the next step in the lecturer's problem solution or answer conceptual questions that emphasize alternative conceptions related to the problem being solved. Formal coaching occurs in laboratory and discussion sections where the students work in assigned mixed ability groups of 3 with 6 groups managed by a teaching assistant. Students work in the same group on context-rich problems in both venues with the primary difference that the problem solution provides the prediction that is tested in the laboratory.

Collaborative groups are used to scaffold more expert-like problem-solving and mitigate the accompanying cognitive load (Sweller, 1988, Johnson & Johnson 2009). Heller and Hollabaugh found that when students work on problem-solving using explicit strategies in mixed ability collaborative groups, they benefit more than when working alone. In this approach, students follow a prescribed problem-solving framework (Polya, 1945, Heller & Reif, 1984) that was modified based on an analysis of the problem solutions of the students. This approach involves both a visualization step and a physics description step where the student creates physics representations and builds a conceptual hierarchy. Students use this framework to solve Context-rich problems while working in groups of 3 with roles that make explicit the metacognitive practices necessary to solve a problem: Manager, Skeptic and Checker/Recorder. They arrived at three as the optimal number of group members by conducting qualitative video research recording various size groups interacting while solving problems (Heller et al, 1992a, Heller and Hollabaugh, 1992b).

They developed context-rich problems, because those are closer to what one may encounter in the real world and also because they are good vehicles for teaching problem-solving: students are more likely to see the benefit of a systematic approach when the problem is challenging. These problems often involve multiple concepts (Badeau et al., 2017, Lin & Singh, 2011b, 2013a, Ibrahim et al., 2017a, 2017b, Ibrahim & Ding, 2021, Yerushalmi et al., 2012a, 2012b) that necessitates building a conceptual hierarchy. For these types of problems, students benefit from a variety of perspectives and can share the cognitive load in a collaborative group (Zhang & Norman 1994, Zhang 1997). To encourage individual learning and discourage free-riding, it is important for students to take on different roles every week and that the groups change members about every 3 weeks. Heller and Hollabaugh found that when using collaborative learning with context-rich problems in the environment of cognitive apprenticeship with the necessary scaffolding (Heller and Heller, 1995), students' problem-solving abilities progress over time and students at all stages of problem-solving development benefit (Heller et al 1992a, Heller and Hollabaugh, 1992b). Other institutions have tested this approach and found it not only substantially increases their student's problem-solving capability but also their conceptual understanding comparable to widely accepted curricular tools especially designed for that purpose (Cummings et al, 1999).

## Integrating Quantitative and Conceptual Understanding in Physics (ICQUIP)

Based on the synergy of conceptual and quantitative knowledge in physics, a framework specific to physics called "Integrating Quantitative and Conceptual Understanding in Physics" (Justice, 2019) or ICQUIP has been proposed. The ICQUIP framework is grounded in the fact that in order to learn physics and develop reasoning skills with quantitative tools, students must be given adequate opportunity to interpret symbolic equations and draw qualitative inferences from them. Without this explicit focus on the integration of conceptual and quantitative aspects of physics learning, quantitative problem-solving can become a mathematical exercise instead of an opportunity to develop reasoning skills and build a robust knowledge structure (Ding et al., 2011). Unfortunately, without explicit guidance and support, many students solve physics problems using superficial clues and cues, and apply concepts essentially by pattern matching. Indeed, many traditional physics courses reward algorithmic problem-solving strategies that many students utilize blindly. Many instructors assume that students know that conceptual analysis and decision making or planning, evaluation of the plan, and reflection on the problem-solving process are as important as the implementation phase of the quantitative problem solution. Some researchers (McDaniel et al., 2016) have found that interactive engagement courses which primarily focus student group work with conceptual questions do not necessarily result in improved problem-solving performance, thus highlighting the importance of integrating both conceptual and quantitative problem-solving when engaging students in interactive activities.

One effective approach consistent with the ICQUIP framework focuses on implementing quantitative problem-solving followed by conceptual problem-solving (Singh, 2008c). These conceptual problems can prompt students to make qualitative inferences from the quantitative problems they just solved. Particular attention is given to designing conceptual problems that probe common misconceptions and challenge students to discriminate between concepts which can easily be confused. It is important that the ICQUIP framework be integrated within the broader cognitive apprenticeship model. In other words, modeling, coaching and scaffolding, and fading help students learn and generalize from integrated conceptual and quantitative problem-solving. Students can be given opportunities to solve coupled quantitative and conceptual problems in small groups where they can be provided appropriate scaffolding. Then in homework the scaffolding can be reduced to help students develop self-reliance.

Prior research suggests that problem-solving that combines quantitative and conceptual problems is an effective instructional strategy for reducing conceptual difficulties and helping students develop a functional understanding of physics (Singh 2008c, Reinhart et al., 2022). For example, in a controlled study in introductory physics, only conceptual problems were posed to some students and quantitative/conceptual problem pairs were posed to others. The students with both problems gave the correct response to both approximately equally and at a much larger rate than the students who were posed only the qualitative problem to theirs (Singh 2008c). The significant improvement in the performance on the conceptual problem for those also working on the quantitative problem and discussions with students suggest that when they were posed both problems, many students recognized their similarity and took advantage of their quantitative solution to solve the conceptual problem.

## Teaching Problem solving in Advanced Physics

Research suggests that many of the issues related to problem-solving in introductory physics are also applicable to advanced physics (Singh & Zhu, 2009), e.g., students must still use effective problem-solving strategies, use problem-solving as an opportunity for learning and organizing their knowledge hierarchically around core concepts and principles. Also, while advanced students may be proficient in calculus used in introductory physics, working with more advanced mathematics can be cognitively demanding while solving physics problems. Therefore, one should carefully consider how to incorporate the mathematics into a course consistent with students' current knowledge so that they do not have cognitive overload while solving problems and they are able to integrate conceptual and quantitative knowledge consistent with the ICQUIP framework. For example, for quantum mechanics, the types of difficulties students have are analogous to those found in introductory physics (Marshman & Singh, 2018).

In one study (Bilak & Singh 2007), introductory students and graduate students performed comparably on conceptual problems involving conductors and insulators. In another study (Maries & Singh, 2013) comparing upper-level undergraduate and graduate students' proficiency with different representations used in electricity and magnetism, both groups had difficulties identifying divergence and curl graphically, although the students were able to calculate them mathematically. These investigations suggest that advanced students may be more proficient with certain types of mathematical manipulations, but many struggle to solve conceptual problems, once again, highlighting the importance of integrating both conceptual and quantitative problem-solving in instruction. Additionally, when it comes to learning from mistakes, similar to introductory students, advanced students also benefit from explicit incentives to correct their mistakes without which they show little improvement on average from midterm to final on similar problems (Brown et al., 2016).

In one study (Lin & Singh, 2010), advanced undergraduate students and quantum physics instructors were asked to categorize quantum mechanics problems based upon similarity of solution. They found while instructors performed well as expected, there was a wide diversity in students' categorization. In another study (Justice, 2019), when upper-level undergraduate students engaged with an integrated qualitative and quantitative tutorial on quantum mechanics and then were tested in a post-test on conceptual problems, they showed similar performance as students who engaged with only a qualitative tutorial. This finding is in contrast to physics graduate students, who performed significantly better on the conceptual problems after engaging with the integrated qualitative and quantitative tutorial (Justice, 2019). One possible reason for this difference between the performance of upper-level undergraduate and graduate students may be the cognitive load while engaging with the integrated tutorial. It is likely that there was a wider diversity in the mathematical preparation of undergraduates so some of them did not benefit as much from the integrated tutorial and may have had cognitive overload while engaging with it. On the other hand, graduate students are likely to have better mathematical preparation and may not have incurred similar cognitive overload. This finding emphasizes the importance of considering cognitive load when designing learning materials for advanced students.
An encouraging research finding pertaining to upper-level undergraduate and graduate students is that in the context of quantum mechanics, they were able to transfer their learning from one context to another on their own without any guidance (Maries et al., 2017, 2020), perhaps suggesting that advanced undergraduate students have developed sufficient metacognitive skills to discern the underlying concepts, and it may be possible to help them transfer their learning from one context to another in other problem-solving situations.

## Consideration of Motivational Factors for Fostering Problem Solving

In any practical classroom environment, it is important to motivate all students to engage in productive problem-solving approaches. We must recognize the diversity of students in our classes and how their background can influence how they perceive the classroom and the extent to which they engage with their peers and materials. Stereotype threat (Wheeler & Petty, 2001, Appel & Kronberger, 2012, Marchand & Taasoobshirazi, 2013) has been shown to negatively impact the performance and engagement in many STEM disciplines of members of the stereotyped groups, e.g., women and racial and ethnic minority students. Issues like lack of sense of belonging and low self-efficacy (belief about one's ability to solve problems) can have a detrimental effect on learning (Marshman et al., 2018b, Binning et al., 2020). Especially for physics problem-solving, students may not believe in their own ability to be successful and this can impact their motivation to engage, and when they do engage in solving problems, anxiety can rob them of their limited working memory (Maloney et al., 2014). Also, students may believe that they cannot become expert problem solvers in the "hard sciences" and there is no point in trying because intelligence is immutable, i.e., they may have a fixed mindset (Dweck, 2007, Yeager & Dweck, 2012). These issues can be particularly important for students from underrepresented groups who do not have role models and are wondering whether they have what it takes to become expert problem solvers. In order for all students to engage meaningfully in problem-solving, it is important for the instructor to create an equitable and inclusive learning environment.

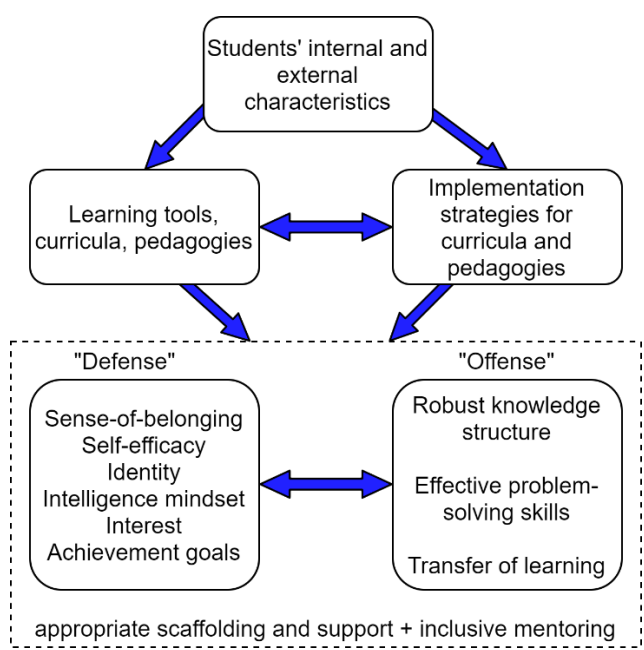

A useful framework to consider in this regard is the Holistic Ecosystem for Learning Physics in an Inclusive and Equitable Environment (HELPIEE) shown in Fig. 3 (Cwik & Singh, 2021). This framework emphasizes that not only should we contemplate students' initial characteristics in the selection and implementation of the learning tools, curricula and pedagogies, but also focus on both defense and offense similar to the coaching in chess or tennis. Here building students' "offense" refers to the types of things instructors often focus on, i.e., helping students learn effective problem-solving approaches, develop the ability to transfer their learning, and develop a robust knowledge structure. "Defense" refers to those issues that physics instructors have generally not considered but which can be particularly detrimental to underrepresented students while learning problem-solving. It refers to increasing students' sense of belonging, self-efficacy and identity as someone who can excel in physics and inculcating a growth mindset, i.e., the fact that the brain is a like muscle and one can develop intelligence by working hard and working smart (Yeager & Dweck, 2012). Also, high physics self-efficacy, identity, and growth mindset have the potential to increase students' physics interest and achievement goals. Unless instructional design endeavors to build up students' "defense", e.g., by creating an equitable and inclusive environment, which can include using interventions at the beginning of class to improve students' sense of belonging, mindset, etc. (Paunesku et al., 2015, Yeager and Walton, 2011 Binning et al., 2020) and then reinforcing the social belonging and mindset messages that everybody is capable of learning physics, many underrepresented students are unlikely to derive the full benefit of pedagogies and curricula designed to foster problem-solving. Additionally, emphasizing that problem-solving is an essential tool for learning physics and that growing one's brain requires struggling with challenging problems and learning from one's mistakes can help promote a growth mindset and improve students' enthusiasm for solving problems.

**Figure 3** Holistic Ecosystem for Learning Physics in an Inclusive and Equitable Environment (HELPIEE) framework.

## Student Attitudes about Problem Solving

Student attitudes about problem-solving can impact how they engage with it. For example, a student who doesn't believe that physics answers should make sense is unlikely to check the answers after solving problems. A student who believes that physics is a collection of disconnected facts and formulas is likely to develop formula-centric approaches to problem-solving. It is therefore important to learn about students' attitudes and approaches to problem-solving and make efforts to improve them while helping students learn physics problem-solving. Related to attitudes about problem-solving are students' attitudes and beliefs about the nature of knowledge in physics (i.e., epistemological beliefs). These include beliefs about whether physics should make intuitive sense, whether it should apply to things in the real world, whether problem-solving involves making sense of the answer, whether learning involves memorizing solutions or understanding the physics, etc. Whenever these are surveyed via instruments like the MPEX (Redish et al., 1998) or C-LASS (Adams et al., 2006) pre- and post-instruction, it is often found that students' attitudes deteriorate, i.e., get farther away from those of a physics expert. However, it is important to recognize that often, the way in which the physics courses are taught can actually contribute to students' attitudes deteriorating. Hammer (Hammer, 1989) tells the story of "Ellen", who was initially motivated to reconcile what she was learning in physics with her intuition and be able to use the concepts to reason about the world, but gradually grew frustrated with the abstract way in which the course was taught and eventually gave up trying to understand the physics and just focused on getting good at identifying the correct formula to use.

Students' attitudes and approaches to problem-solving vary from introductory to advanced courses (Cummings et al., 2004, Mason & Singh, 2010b). For example, in response to the statement "I always identify the physics principles involved in the problem first before looking for corresponding equations," 62% of introductory students responded favorably. While one can look at this positively because more than half of the students have the expert-like attitude, it does mean that 38% of the students are taking a course in which they do not perceive identifying the principle as important. Similarly, 55% of students agree with the statement: "When I solve physics/astronomy problems, I always explicitly think about the concepts that underlie the problem." These questions relate to a factor described as "metacognition and enjoyment of problem-solving". In one investigation (Good et al., 2019) the attitudes and approaches to problem-solving of male and female students in introductory courses were compared. It was found that female students often valued better approaches to problem-solving and their beliefs were unchanged from the beginning to the end of the course as opposed to those of male students that deteriorated. Interviews suggest that on items such as "When solving physics problems, I often find it useful to first draw a picture or diagram of the situations described in the problems", female students often agreed with the statements and noted that they always draw diagrams because otherwise they did not think they would be able to solve the physics problem correctly (Good et al., 2019). Interviews also hint at the fact that the lower self-efficacy of women about physics problem-solving may partly be responsible for their better approaches to problem-solving. However, it would be valuable for instructors to take advantage of female students' good approaches to problem-solving to help them become proficient problem solvers. Good et al. (Good et al., 2018b) also compared the attitudes and approaches to problem-solving of students in introductory physics and astronomy courses. They found that astronomy students had better attitudes towards problem-solving. Interviews suggest that the more interesting and engaging contexts of astronomy were at least partly responsible for this finding. This study suggests that instructors should consider using more interesting and engaging contexts in physics problems including those that are more meaningful to students (as opposed to choosing a context such as a box on an inclined plane).

## Professional development of TAs and instructors: Beliefs about problem-solving

To make a greater impact on the teaching of effective problem-solving in physics, it is important to engage in the professional development of the educators of tomorrow (Maries, 2020) and professional development for the professors of today. In order for these programs to be successful, it is important to be aware of the conceptions related to problem-solving of physics instructors and teaching assistants.

The good news is that physics instructors at R1 universities believe that problem-solving is a central goal of a physics course and that solving problems requires conceptual understanding (Yerushalmi et al., 2007, 2010, Henderson et al., 2007). In interviews with six faculty (Yerushalmi et al., 2007), researchers found that instructors have relatively low expectations of "typical" students who they felt often use poor problem-solving strategies. There is an apparent contradiction at play because while instructors value problem-solving and believe that it is important, they seem to believe that few introductory students are capable of learning it. Another apparent contradiction is that while they expect students to be reflective in their learning from problem-solving, their instruction was not typically designed for reflection and they didn't recognize their own role in helping students develop their metacognitive skills. They felt that students should do it on their own. Along similar lines, a study by Hora (Hora, 2014) investigated the beliefs of 56 math and science instructors at undergraduate universities. The research revealed several instructor beliefs about student learning in general, and in the context of solving problems. These included that students learn best by practice and perseverance, by articulating their own thoughts, ideas, and problem-solving processes to others, and that learning is best facilitated by active, hands-on engagement with the material.

However, while some instructors have productive ideas when it comes to problem-solving (Yerushalmi et al., 2007, 2010, Hora, 2014), when it comes to grading physics problems, their ideas may not be as productive. Henderson et al. (Henderson et al., 2004) found that when grading, instructors may sometimes engage in practices that are detrimental to students' learning of problem-solving. In particular, when grading example solutions, instructors sometimes rewarded brief solutions and placed the burden of proof on themselves when grading, not encouraging students to show their work. Their reluctance to take off points because nothing was incorrect in a brief, three-line student solution to a problem, but their willingness to take off points due to canceling mistakes in another student solution that included showing complete steps and use of effective problem-solving strategies sends the wrong message to students. In other words, from getting penalized for making mistakes when showing their work, students may learn that the best types of solutions are the ones in which little is shown so as to avoid the chance of getting penalized.

The TAs show similar reluctance (Marshman et al., 2018a, Marshman et al., 2020b) when it comes to grading, and since they are most commonly responsible for grading student work, their grading practices may be counterproductive to student learning of problem-solving. It's also important to note that TAs may view problem-solving differently in introductory physics compared to advanced physics. For example, Marshman et al. (Marshman et al., 2017) found that graduate TAs had different expectations of introductory and advanced students and were likely to grade introductory students much more leniently because they did not think they were capable of using effective problem-solving strategies in solving physics problems. In particular, the TAs felt that being systematic in problem-solving is more important in advanced physics, but it is not necessarily required of introductory students and it may also be beyond their capabilities. Thus, it is important to be aware of research on graduate TAs' conceptions with regard to problem-solving. Since graduate TAs were recently students in introductory physics classrooms, one might expect TAs' beliefs regarding teaching problem-solving to be influenced by their previous physics instructors.

Furthermore, TAs' beliefs about teaching and learning may be influenced by their own experiences as learners and the type of instruction that was beneficial for them. For example, in the laboratory context, TAs believe that students learn similar to how they learned and implement instructional strategies that were effective for them, but not necessarily beneficial for students (Seung, 2012). Moreover, TAs acknowledge instructional strategies from educational research, but sometimes disregard them in favor of their own views of what is appropriate instruction, e.g., that they should explain to students how to solve problems by working out problems carefully on the board (Seung, 2012). In a similar vein, TAs' own approaches to problem-solving can serve as indicators of their beliefs regarding learning and teaching problem-solving. Furthermore, Mason and Singh (Mason & Singh, 2010b) found that while nearly 90% of graduate students reported that they explicitly think about the underlying concepts when solving introductory physics problems, approximately 30% of them stated that solving introductory physics problems merely requires a "plug and chug" strategy. Interviews with graduate students suggest that when they solve introductory problems, they turn out to be exercises as opposed to problems for them (i.e., the solution is "obvious" to graduate students and they do not need to use effective problem-solving approaches that would otherwise be essential for introductory students). Since the TAs can immediately recognize the principles required to solve the introductory problem, they perceive introductory problem-solving as not requiring much thought or reflection. However, thinking about the difficulty of the problems from their own perspective instead of introductory students' perspective can be detrimental to helping introductory students learn effective problem-solving. These findings suggest that TAs who teach recitations or laboratory sections may not model, coach, or assess explication of reasoning or reflection because they were not necessary for them to solve introductory physics problems, although they are highly beneficial to introductory physics students.

Lin et al. (Lin et al., 2013) studied TAs' beliefs about learning and teaching of problem-solving using example problem solutions. This study revealed a discrepancy between TAs' stated goals and practices. For example, when TAs were asked to evaluate three different versions of example solutions, many valued solutions that include features which were supportive of helping students develop effective problem-solving approaches. Most TAs expressed process-oriented learning goals (i.e., helping students become more systematic in their problem-solving approaches and make better use of problem-solving as a tool for learning) when contemplating the use of example solutions in introductory physics. However, their own designed example solutions did not include features supportive of helping students learn effective problem-solving approaches. When TAs were unaware of the conflict between their stated goals and practices, they tended to prefer product-oriented solutions (i.e., solutions in which the reasoning is not explicated). A similar discrepancy may arise in the context of grading, i.e., TAs may have productive beliefs about the role of grading in the learning process, but employ grading practices which do not align with those beliefs, and they may not even be aware of the discrepancy.

Additionally, with regards to teaching problem-solving, prior research show that TAs prefer using problems that are broken into parts to context-rich problems (Good et al., 2018a, 2020). While problems that are broken into parts are good initially to provide scaffolding for students (Collins et al., 1989), that scaffolding support should be removed to help students develop self-reliance. In fact, the TAs saw little to no use for the context-rich problems, even in contexts of students working collaboratively. It is also important to recognize the extent to which TAs consider student evaluations in their decisions related to teaching. For example, TAs may want to foster reasoning and problem-solving, but are afraid that students wouldn't like it and would complain. Therefore, they do something that is in between what they want to do and what they think students would like (Chini & Al-Rawi, 2012). However, there is hope. TAs respond to a well-structured orientation and support system that emphasizes their importance for student learning and lets them discuss and experience research-based instruction (Lawrenz et al, 1992). It is especially important that they are put in a teaching role compatible with their teaching and physics knowledge where they will succeed and become the next generation of teaching ready faculty.

## Other Related Frameworks

There are a few other frameworks that are closely related to problem-solving. We have not included these earlier because they have traditionally not been referred to as problem-solving research (although they cite problem-solving research). For instance, the epistemic games framework (Tuminaro & Redish, 2007, Hu et al., 2019) relates to what students do when engaged in problem-solving. The framework tries to make sense of how students use physics and mathematics concepts when problem-solving and how they blend these. Some of the games identified relate to specific types of strategies used by experts, e.g., using a diagram in a qualitative analysis, which is similar to what Tuminaro and Redish refer to as the "Pictorial analysis game". While some of the epistemic games research investigates how students blend mathematics with physics, other epistemic games research refers to sense-making and meaning-making, or blending of conceptual and quantitative, of physics and math, etc. (Eichenlaub & Redish, 2019). Moreover, the dual process theory framework (Kahneman, 2013, Kryjevskaia et al., 2021)) focuses on fast (reflexive) and slow (reflective) thought processes while solving problems. In problem-solving research, experts have the ability to be reflective and use effective problem-solving strategies but novices are not often reflective while solving problems and need scaffolding support in learning to be reflective.

It is also interesting to recognize that many of the effective approaches associated with problem-solving discussed here are important for students to learn because they are training to become scientists themselves and those approaches would also help them solve authentic scientific problems (Price et al., 2021, Leak et al., 2017). For example, Price et al. (Price et al., 2021) interviewed 52 successful scientists and engineers from a wide range of disciplines about the problem-solving behaviors they use when solving authentic problems in their discipline and developed a decision framework involving 29 distinct decisions that were common across science and engineering disciplines. Many of these decisions closely relate to what has been discussed here, e.g., framing and planning the problem, interpreting and choosing solutions, and reflection are part of the framework. Similar to the discussion here, they advocate using the deliberate practice model within a broader cognitive apprenticeship model and suggest that instructors should provide training opportunities for students to engage in making the 29 decisions from their framework. They also point out that since the decisions were obtained from experts discussing how they solved authentic problems in their disciplines, when teaching problem-solving, only a subset of those decisions would be relevant for a particular task that students may be asked to engage in.

## Conclusion

Helping students learn effective problem-solving strategies is recognized as an explicit goal in many physics courses for science and engineering majors. It is important that research-validated approaches to fostering problem-solving are incorporated explicitly in instruction. Being aware of all the different aspects of effective problem-solving can help guide instructional design. Many tools have been developed and the extent to which students will engage with them meaningfully depends on how they are incentivized and incorporated in the learning objectives of the course, its instructional design as well as assessment. To underscore the impact that research on problem-solving has had in teaching, in the last decade, many book publishers attempt to incorporate some of the research-based methods in their most popular books, and although it is unclear the extent to which this has actually helped improve student learning, it is an encouraging development nevertheless.

# References

Adams, W. et al. (2006). New instrument for measuring student beliefs about physics and learning physics: The Colorado Learning Attitudes about Science Survey. *Phys. Rev. PER, 2,* 010101.

Anderson, J. (1995). *Learning and Memory*. New York: Wiley.

Anderson, J. (2000). Problem solving. In J. Anderson (Ed.). *Cognitive Psychology and its Implications* (pp. 239–278). New York: Worth.

Appel, M., & Kronberger, N. (2012). Stereotypes and the achievement gap: Stereotype threat prior to test taking. *Educ. Psych. Rev., 24*, 609.

Badeau, R. et al. (2017). What works with worked examples: Extending self-explanation and analogical comparison to synthesis problems. *Phys. Rev. PER*, *13*, 020112.

Bagno, E., & Eylon, B. (1997). From problem solving to a knowledge structure: An example from the domain of electromagnetism. *Am. J. Phys., 65*(8), 726-736.

Bagno, E. et al. (2000). From fragmented knowledge to a knowledge structure: Linking the domains of mechanics and electromagnetism. *Am. J. Phys., 67*(7), S16-S26.

Bassok, M., & Holyoak, K. (1989). Interdomain transfer between isomorphic topics in algebra and physics. *J. Exp. Psych.: Learning Memory and Cognition, 15*(1), 153-166.

Beichner, R. (1999). "Goal oriented problem solving". retrieved from https://projects.ncsu.edu/per/archive/GOALPaper.pdf

Beilock, S., McConnell, A., Rydell, R. (2007). Stereotype threat and working memory: Mechanisms, alleviation, and spillover. *J. Exp. Psych.: Gen., 136*(2), 256-276.

Beilock, S. et al. (2006). On the causal mechanisms of stereotype threat: Can skills that don't rely heavily on working memory still be threatened? *Pers. Soc. Psych. Bull., 38,* 1059-1071.

Bilak, J., & Singh, C. (2007). Improving students' conceptual understanding of conductors and insulators. *AIP Conf. Proc., 951,* 49-52.

Binning, K. et al. (2020). Changing social contexts to foster equity in college science courses: An ecological-belonging intervention. *Psychol. Sci., 31*(9), 1059-1070.

Bransford, J. et al (2000) How People Learn: Brain, Mind, Experience, and School: Expanded Edition. The National Academies Press.

Brown, B. et al. (2016). Improving performance in quantum mechanics with explicit incentives to correct mistakes. *Phys. Rev. PER, 12,* 010121.

Caballero, M., Engelhardt L., Knaub A., Kuchera M., Lopez del Puerto,M., Lunk, B., Roos, K., Zimmerman, T. (2021), PICUP Virtual Capstone Conference Report, Hilborn, R. ed, Partnership for the Integration of Computation in Undergraduate Physics, https://www.compadre.org/picup//events/pdfs/2021_PICUP_Capstone_Report_Final_Final_220502.pdf.

Chase, W., & Simon, H. (1973). Perception in chess. *Cog. Psych., 4*, 55-81.

Chi, M. et al. (1981). Categorization and representation of physics knowledge by experts and novices. *Cog. Sci., 5*, 121-151.

Chi, M., & Glaser, R. (1985). Problem solving ability. In R.J. Sternberg (Ed.). *Human Abilities: An Information Processing Approach* (pp. 227–250). New York: Freeman.

Chini, J., & Al-Rawi, A. (2012). Alignment of TAs' beliefs with practice and student perception. *AIP Conf. Proc., 1513,* 98-101.

Chipman, S. et al. (2000). Introduction to cognitive task analysis. In J. Schraagen, S. Chipman & V. Shute (Eds.). *Cognitive Task Analysis* (pp. 3-23). Mahwah, New Jersey: Lawrence Erlbaum Associates.

Clement, J. (1998). Observed methods for generating analogies in scientific problem solving. *Cog. Sci., 12*(4), 563-586.

Collins, A. et al. (1989). Cognitive apprenticeship: Teaching the crafts of reading, writing and apprenticeship. In R. Glaser and L. Resnick (Eds.). *Knowing, Learning and Instruction: Essays in Honor of Robert Glaser* (pp. 453-494). Hillsdale, New Jersey: Lawrence Erlbaum Associates.

Crane, H. (1966). Experiments in Teaching Captives. American Journal of Physics 34, 799-807.

Crane, H. (1969). Problems for Introductory Physics. The Physics Teacher 7, 371-378.

Cummings, K. et al. (1999). Evaluating innovation in studio physics. American Journal of Physics 67, 38-44.

Cummings, K. et al. (2004). Attitudes toward problem solving as predictors of student success. *AIP Conf. Proc., 720*, 133-136.

Cwik, S., & Singh, C. (2023). Framework for and Review of Research on Assessing and Improving Equity and Inclusion in Undergraduate Physics Learning Environments. *Handbook of Research on Physics Education, Volume 3: Physics Education Research Special Topics*, edited by M. F. Taşar and P. Heron (AIP Publishing, Melville, New York, 2023), p. 2-1.

de Jong, T., & Ferguson-Hessler, G. (1986). Cognitive structures of good and poor novice problem solvers in physics. *J. Educ. Psych., 78*(4) 279-288.

DeVore, S. et al. (2017). Challenge of engaging all students via self-paced interactive electronic learning tutorials for introductory physics. *Phys. Rev. PER, 13*, 010127.

Dewey, J.(1910) How We Think, D.C. Heath & Co.

Ding, L. et al. (2011). Exploring the role of conceptual scaffolding in solving synthesis problems. *Phys. Rev. ST-PER, 7,* 020109.

Docktor, J. et al. (2012). Impact of a short intervention on novices' categorization criteria. *Phys. Rev. ST-PER, 8,* 020102.

Docktor, J. et al. (2015). Conceptual problem solving in high school physics. *Phys. Rev. ST-PER, 11*, 020106.

Dorst, K. (2015). Frame Creation and Design in the Expanded Field, *She Ji: The Journal of Design, Economics, and Innovation 1*(1), 22-33.

Dreyfus, S. (2004). The Five-Stage Model of Adult Skill Acquisition, Bulletin of Science, Technology & Society, 24, 177-181

Dufresne, R. et al. (1997). Solving physics problems with multiple representations. *Phys. Teach., 35*(5), 270-275.

Dufresne, R. et al. (1992). Constraining novices to perform expertlike problem analyses: Effects on schema acquisition. *J. Learn. Sci., 2*(3), 307-331.

Dweck, C. (2007). Is math a gift? Beliefs that put females at risk, In S. J. Ceci and W. M. Williams (Eds.). *Why Aren't More Women in Science? Top Researchers Debate the Evidence* (pp. 47–55). American Psychological Association.

Eichenlaub, M., & Redish, E. (2019). Blending Physical Knowledge with Mathematical Form in Physics Problem Solving. In G. Pospiech, M. Michelini, & B.-S. Eylon (Eds.). *Mathematics in Physics Education* (pp. 127-151). Cham, Switzerland: Springer.

Ericsson, K. et al. (1993). The role of deliberate practice in the acquisition of expert performance. *Psychol. Rev., 100*(3), 363–406.

Eylon, B., & Reif, F. (1984). Effects of knowledge organization on task performance. *Cog. Instruct., 1*(1), 5-44.

Flavell, J. (1976) Metacognition and cognitive monitoring: a new area of cognitive developmental inquiry., American Psychologist, 34.,906-911

Foley A, Herts J, Borgonovi F, Guerriero S, Levine S, Beilock S (2017) The Math Anxiety-Performance Link: A Global Phenomenon. *Current Directions in Psychological Science*. 26(1),52-58.

Funahashi, S. (2017) <u>Working memory in the prefrontal cortex</u>**,** Brain sciences, 7, 49-71

Good, M. et al. (2018a). Physics teaching assistants' views of different types of introductory problems: Challenge of perceiving the instructional benefits of context-rich and multiple-choice problems. *Phys. Rev. PER, 14,* 020120.

Good, M. et al. (2018b). Comparing introductory physics and astronomy students' attitudes and approaches to problem solving. *Eur. J. Phys., 39,* 065702.

Good, M. et al. (2019). Impact of traditional or evidence-based active-engagement instruction on introductory female and male students' attitudes and approaches to physics problem solving. *Phys. Rev. PER, 15,* 020129.

Good, M. et al. (2020). Graduate teaching assistants' views of broken-into-parts physics problems: Preference for guidance overshadows development of self-reliance in problem solving. *Phys. Rev. PER, 16,* 010128.

Hammer, D., (1989). Two approaches to learning physics. *Phys. Teach., 27*(9) 664-670.

Hardiman, P. et al. (1989). The relationship between problem categorization and problem solving among experts and novices. *Mem. Cogn. 17,* 627–638.

Hebb, D. (1949) The Organization of Behavior, John Wiley and Sons

Heller, K. & Heller, P. (1995). The competent problem solver, a strategy for solving problems in physics (2nd ed.). McGraw-Hill.

Heller, K. & Heller, P. (2010). Cooperative problem solving: a user's manual. University of Minnesota, https://www.aapt.org/conferences/newfaculty/upload/coop-problem-solving-guide.pdf.

Heller, P., Keith, R. & Anderson, S. (1992a). Teaching problem solving through cooperative grouping. Part 1. Group versus individual problem solving. *Am. J. Phys., 60*, 627.

Heller, P., & Hollabaugh, M. (1992b). Teaching problem solving through cooperative grouping. Part 2. Designing problems and structuring groups. *Am. J. Phys., 60*, 637.

Heller, J., & Reif, F. (1982). Knowledge structure and problem solving in physics. *Educ. Psych., 17*(2), 102-127.

Heller, J., & Reif, F. (1984). Prescribing effective human problem-solving processes: Problem description in physics. *Cog. Instruct., 1*(2), 177-216.

Henderson, C. et al. (2004). Grading student problem solutions: The challenge of sending a consistent message. *Am. J. Phys., 72*, 164-169.

Henderson, C. et al. (2007). Physics faculty beliefs and values about the teaching and learning of problem solving. II. Procedures for measurement and analysis. *Phys. Rev. ST-PER, 3*, 020110.

Holyoak, K. (1985). The pragmatics of analogical transfer. In G. H. Bower (Ed.). *The Psychology of Learning and Motivation* (pp. 59-87). New York: Academic Press.

Hora, M. (2014). Exploring faculty beliefs about student learning and their role in instructional decision-making. *Rev. High. Educ., 38* (1), 37-70.

Huffman, D. (1997). Effect of explicit problem solving strategies on high school students' problem-solving performance and conceptual understanding of physics. *J. Res. Sci. Teach., 34*(6), 551-570.

Hsu, L., & Heller, K. (2005). Computer problem solving coaches. *AIP Conf. Proc., 790, 197-201.*

Hu, D. et al. (2019). Characterizing mathematical problem solving in physics-related workplaces using epistemic games. *Phys. Rev. PER, 15*, 020131.

Ibrahim, B., & Ding, L. (2021). Sequential and simultaneous synthesis problem solving: A comparison of students' gaze transitions, *Phys. Rev. PER, 17*, 010126.

Ibrahim, B. et al. (2017a). Students' conceptual performance on synthesis physics problems with varying mathematical complexity. *Phys. Rev. PER, 13*, 010133.

Ibrahim, B. et al. (2017b). How students process equations in solving quantitative synthesis problems? Role of mathematical complexity in students' mathematical performance. *Phys. Rev. PER, 13*, 020120.

Jacobs, J., & Paris, S. (1987). Children's metacognition about reading. Issues in definition, measurement, and instruction. *Educ. Psych., 22*, 255-278.

Johnson, D., & Johnson, R. (2009). An educational psychology success story: Social interdependence theory and cooperative learning. *Educ. Res., 38*(5), 365-379.

Justice, P. (2019). Helping Students Learn Quantum Mechanics using Research-Validated Learning Tools. PhD Dissertation. University of Pittsburgh.

Lawrenz, F., Heller, P., Keith, R., Heller, K. (1992) Training the Teaching Assistant: Matching TA Strengths and Capabilities to Meet Specific Program Goals, *Journal of College Science Teaching* 22 (2), 106-109.

Kahneman, D. (2013). *Thinking Fast and Slow.* (Farrar Strauss and Giroux, New York, NY).

Kail, R., & Salthouse, T. (1994). Processing speed as a mental capacity. *Acta Psychologica, 86*, 199-225.

Koedinger, K. R., & Nathan, M. J. (2009). The Real Story Behind Story Problems: Effects of Representations on Quantitative Reasoning. *Journal of the Learning Sciences, 13*(2), 129-164.

Koenig, K. et al. (2022). Promoting problem solving through interactive video-enhanced tutorials. *Phys. Teach., 60*, 331-334.

Kryjevskaia, M. et al. (2021). Intuitive or rational? Students and experts need to be both. *Phys. Teach., 8*, 28-31.

Kyllonen, P., & Christal, R. (1990). Reasoning ability is (little more than) working memory capacity?! *Intelligence, 14*, 389-433.

Larkin, J., & Simon, H. (1987). Why a diagram is (sometimes) worth ten thousand words. *Cog. Sci., 11*(1), 65-99 (1987).

Larkin, J. et al. (1980). Expert and novice performance in solving problems. *Science, 208*, 1335.

Leak, A. et al. (2017). Examining problem solving in physics-intensive Ph.D. research. *Phys. Rev. PER, 13*, 020101.

Lin, S.-Y., & Singh, C. (2010). Categorization of quantum mechanics problems by professors and students. *Eur. J. Phys., 31*, 57-68 (2010).

Lin, S.-Y., & Singh, C. (2011a). Challenges in using analogies. *Phys. Teach., 49*(8), 512-513.

Lin, S.-Y., & Singh, C. (2011b). Using isomorphic problems to learn introductory physics. *Phys. Rev. ST-PER, 7*(2), 020104.

Lin, S.-Y., & Singh, C. (2013a). Using an isomorphic problem pair to learn introductory physics: Transferring from a two-step problem to a three-step problem. *Phys. Rev. ST-PER, 9*, 020114.

Lin, S.-Y., & Singh, C. (2013b). Effect of scaffolding on helping introductory physics students solve quantitative problems involving strong alternative conceptions. *Phys. Rev. ST-PER, 11*, 020105.

Lin, S.-Y., Henderson, C., Mamudi, W., Singh, C., & Yerushalmi, E. (2013). Teaching assistants' beliefs regarding example solutions in introductory physics. *Physical Review Special Topics: Physics Education Research, 9*, 010120 (2013).

Maloney, E. et al. (2014). Anxiety and Cognition. *WIREs Cognitive Science, 5,* 403.

Marchand, G., & Taasoobshirazi, G. (2013). Stereotype threat and women's performance in physics. *Int. J. Sci. Educ, 35,* 3050.

Maries, A. (2020). Preparing the next generation for active learning. In J. Mintzes and E. Walter (Eds.). *Active learning in college science: The case for evidence based practice* (pp. 965-983). Berlin: Springer Nature.

Maries, A., & Singh, C. (2013). Core graduate courses: A missed learning opportunity? *AIP. Conf. Proc., 1513*, 382-385.

Maries, A., & Singh, C. (2018a). Do students benefit from drawing productive diagrams themselves while solving introductory physics problems? The case of two electrostatics problems. *Eur. J. Phys., 39*, 015703.

Maries, A., & Singh, C. (2018b). Case of two electrostatics problems: Can providing a diagram adversely impact introductory physics students' problem solving performance? *Phys. Rev. PER, 14*, 010114.

Maries, A. et al. (2017). Effectiveness of interactive tutorials in promoting "which-path" information reasoning in advanced quantum mechanics. *Phys. Rev. PER, 13*, 020115.

Maries, A. et al. (2020). Can students apply the concept of "which-path" information learned in the context of Mach-Zehnder Interferometer to the double slit experiment? *Am. J. Phys., 88* (7), 542-550.

Marshman, E., & Singh, C. (2018). Framework for understanding the patterns of student difficulties in quantum mechanics. *Phys. Rev. PER, 11*, 020119.

Marshman, E. et al. (2017). Contrasting grading approaches in introductory physics and quantum mechanics: The case of graduate teaching assistants. *Phys. Rev. PER, 13*, 010120.

Marshman, E. et al. (2018a). The challenges of changing teaching assistants' grading practices: Requiring students to show evidence of understanding. *Can. J. Phys., 96* (4), 420-437.

Marshman, E. et al. (2018b). Female students with A's have similar physics self-efficacy as male students with C's in introductory courses: A cause for alarm? *Phys. Rev. PER, 14*, 020123 (2018).

Marshman, E. et al. (2020a). Holistic framework to help students learn effectively from research-validated self-paced learning tools. *Phys. Rev. PER, 16*, 020108.

Marshman, E. et al. (2020b). Physics postgraduate teaching assistants' grading approaches: Conflicting goals and practices. *Eur. J. Phys., 41*, 055701.

Martinez, M. (1998) What is Problem Solving?, Phi Delta Kappan 79, 605-609,

Martinez, M. (2006) What is Metacognition? Phi Delta Kappan 87, 696-699,

Mason, A., & Singh, C. (2010a). Helping students learn effective problem solving strategies by reflecting with peers. *Am. J. Phys., 78*(7), 748-754.

Mason, A., & Singh, C. (2010b). Surveying graduate students' attitudes and approaches to problem solving. *Phys. Rev. ST-PER, 6*, 020124.

Mason, A., & Singh, C. (2011). Assessing expertise in introductory physics using categorization task. *Phys. Rev. ST-PER, 7*, 020110.

Mason, A., & Singh, C. (2016a). Impact of guided reflection with peers on the development of effective problem solving strategies and physics learning. *Phys. Teach., 54*, 295-299.

Mason, A., & Singh, C. (2016b). Using categorization of problems as an instructional tool to help introductory students learn physics, *Phys. Educ., 51*, 025009.

Mazur, E. (1997). *Peer Instruction: A User's Manual.* Englewood Cliffs, Prentice-Hall.

McDaniel, M. et al. (2016). Dissociative conceptual and quantitative problem solving outcomes across interactive engagement and traditional format introductory physics. *Phys. Rev. PER, 12,* 020141.

McDermott, L. (1991). What we teach and what is learned—Closing the gap, *Am. J. Phys* 59, 301.

McDermott, L. (2001). Oersted Medal Lecture 2001: Physics education research–The key to student learning. *Am. J. Phys., 69,* 1127.

Miller, G. (1956). The magical number seven, plus or minus two: Some limits on our capacity for processing information. *Psychol. Rev., 63*(2), 81-97.

Morphew, J. et al. (2020). Effect of presentation style and problem-solving attempts on metacognition and learning from solution videos. *Phys. Rev. PER, 16*, 010104.

Nathan, M. et al. (2001). Expert blind spot: When content knowledge eclipses pedagogical content knowledge. In L. Chen et al. (Eds.). *Proceeding of the Third International Conference on Cognitive Science* (pp. 644-648). Beijing, China: USTC Press.

Newell, A. et al. (1958). Elements of a theory of human problem solving. *Psychol. Rev., 65*(3), 151.

Nguyen, D.-H. et al. (2010). Facilitating students' problem solving across multiple representations in introductory mechanics. *AIP Conf. Proc., 1289*, 45-48.

Novick, L. (1988). Analogical transfer, problem similarity, and expertise. *J. Exp. Psych.: Learning, Memory, and Cognition, 14*(3), 510–520.

Oerlein, K. (1937). Mathematical Difficulty in College Physics. The Mathematics Teacher, 30 (3), 125-127

Palincsar, A., & Brown, A. (1983). Reciprocal teaching of comprehension-monitoring activities, *Technical Report No. 269*, Center for the Study of Reading.

Paunesku, D. et al. (2015). Mind-set interventions are a scaleable treatment for academic underachievement. *Psychol. Sci., 26*, 784.

Piaget, J. (1967) Six psychological studies, Random House

Polya (1945). *How to Solve It.* Princeton, New Jersey: Princeton University Press.

Price, A. et al. (2021). A detailed characterization of the expert problem-solving process in science and engineering: Guidance for teaching and assessment. *CBE-Life Sci. Educ., 20*, 1-15.

Quarantotto, D. (2020) Aristotle on Science as Problem Solving. Topoi. 39, 857-868

Redish, et al. (1998). Student expectations in introductory physics. *Am. J. Phys., 66*(3), 212-224.

Rees, P. et al (2016) The emergence of neuroscientific evidence on brain plasticity: Implications for educational practice, Educational & Child Psychology 33, 8-19

Reif, F. (1995). Millikan lecture 1994: Understanding and teaching important scientific thought processes. *Am. J. Phys., 63*(1), 17-32.

Reif, F., & Scott, L. (1999). Teaching scientific thinking skills: Students and computers coaching each other. *Am. J. Phys., 67* (9) 819-831.

Reinhard, A. et al. (2022). Assessing the impact of metacognitive postreflection exercises on problem-solving skillfulness. *Phys. Rev. PER, 18*, 010109.

Rosengrant, D. et al. (2005). Free-body diagrams - Necessary or sufficient. *AIP Conf. Proc., 790*, 177.

Ryan, Q. et al. (2016). Computer problem-solving coaches for introductory physics: Design and usability studies. *Phys. Rev. PER, 12*, 010105.

Schoenfeld. A. (1987). What's all the fuss about metacognition. In A. Schoenfeld (Ed.). *Cognitive Science and Mathematics Education* (pp. 189-215). Hillsdale, New Jersey: Lawrence Erlbaum Associates.

Schoenfeld, A. & Herman., D. (1982). Problem perception and knowledge structure in expert and novice problem solvers. *J. Exp. Psych.: Learning, Memory and Cognition 8*(5), 484-494.

Schraw, G. (1998). Promoting general metacognitive awareness. *Instruct. Sci., 26,* 113–125.

Seung. E., (2012). The process of physics teaching assistants' pedagogical content knowledge development. *Int. J. Sci. Math. Educ., 11*, 1303.

Shuell, T (1990) Teaching and learning as problem solving, Theory Into Practice, 29, 102-108)

Simon, H. (1974). How big is a chunk?. *Science, 183*(4124), 482-488.

Simon, H. & Simon, D. (1978). Individual differences in solving physics problems. In R. S. Siegler (Ed.), Children's thinking: What develops? (pp. 325–348). Lawrence Erlbaum Associates, Inc.

Singh, C. (2002). When physical intuition fails. *Am. J. Phys., 70*, 1103-1109.

Singh, C. (2008a). Assessing student expertise in introductory physics with isomorphic problems, Part I: Performance on a non-intuitive problem pair from introductory physics. *Phys. Rev. ST-PER, 4*, 010104.

Singh, C. (2008b). Assessing student expertise in introductory physics with isomorphic problems, Part II: Examining the effect of some potential factors on problem solving and transfer. *Phys. Rev. ST-PER, 4*, 010105.

Singh, C. (2008c). Coupling conceptual and quantitative problems to develop student expertise in introductory physics. *AIP Conf. Proc., 1064*, 199-202, (2008).

Singh, C. (2009a). Categorization of problems to assess and improve proficiency as teachers and learners. *Am. J. Phys., 77*(1) 73-80.

Singh, C. (2009b). Problem solving and learning. *AIP Conf. Proc., 1140*, 183-197.

Singh, C., & Zhu, G. (2009). Cognitive issues in learning advanced physics: An example from quantum mechanics. *AIP Conf. Proc., 1179*, 63-66.

Sweller, J. (1988). Cognitive load during problem solving: Effects on learning. *Cog. Sci., 12*(2), 257-285.

Sweller, J. (2011). Cognitive Load Theory. In J. Mestre & B Ross (Eds.). *The Psychology of Learning and Motivation* (pp. 37-76). San Diego, California: Elsevier Academic Press.

Sweller, J. et al. (2019). Cognitive architecture and instructional design: 20 years later. *Educ. Psych. Rev., 31*, 261-292.

Tuminaro, J., & Redish, E. (2007). Elements of a cognitive model of physics problem solving: Epistemic games. *Phys. Rev. ST-PER, 3*, 020101 (2007).

Turing, A. (1950) Can a Machine Think? Mind, Vol 59, 433–460

Van Heuvelen, A. (1991a). Overview, case study physics. *Am. J. Phys., 59*(10), 898-907.

Van Heuvelen, A. (1991b). Learning to think like a physicist: A review of research-based instructional strategies. *Am. J. Phys., 59*(10), 891-897.

Van Heuvelen, A., & Zou, X. (2001). Multiple representations of work-energy processes. *Am. J. Phys., 69*(2), 184-194.

van Someren, M. W., Reimann, P., Boshuizen, H. P. A., & de Jong, T. (1998). *Learning with Multiple Representations.* New York, NY, Elsevier Science, Inc.

Vygotsky, L. (1978). *Mind in Society: The Development of Higher Psychological Processes.* Cambridge, Massachusetts: Harvard University Press.

Warnakulasooriya, R. et al. (2007). Time to completion of web-based physics problems with tutoring. *Journal of the Experimental Analysis of Behavior, 88*, 103-113.

Wheeler, S., & Petty, R. (2001). The effects of stereotype activation on behavior: A review of possible mechanisms, *Psychol. Bull., 127*, 797.

Whitcomb, K. et al. (2021). Improving accuracy in measuring the impact of online instruction on students' ability to transfer physics problem-solving skills. *Phys. Rev. ST-PER, 17*, 010112.

Wiggins, G., & McTighe, J. (2005). *Understanding by design, 2nd ed*. (Association for Supervision and Curriculum Development ASCD, Alexandria, VA).

Wood, D., Bruner, J., & Ross, G. (1976). The role of tutoring in problem solving. Journal of Child Psychology and Psychiatry and Allied Disciplines, 17(2), 89–100

Yeager, D., & Dweck, C. (2012). Mindsets that promote resilience: When students believe that personal characteristics can be developed. *Educ. Psych., 47*, 302 (2012).

Yeager, D., & Walton, G. (2011). Social-psychological interventions in education: They're not magic. *Rev. Educ. Res., 81*(2), 267–301.

Yerushalmi, E. et al. (2007). Physics faculty beliefs and values about the teaching and learning of problem solving. I. Mapping the common core. *Phys. Rev. ST-PER, 3,* 020109.

Yerushalmi, E. et al. (2012a). What do students do when asked to diagnose their mistakes? Does it help them? I. An atypical quiz context. *Phys. Rev. ST-PER, 8,* 020109 (2012).

Yerushalmi, E. et al. (2012b). What do students do when asked to diagnose their mistakes? Does it help them? II. A more typical quiz context. *Phys. Rev. ST-PER, 8,* 020110 (2012).

Yerushalmi, E. et al. (2010). Instructors' reasons for choosing problem features in a calculus-based introductory physics course. *Phys. Rev. ST-PER, 6,* 020108.

Zhang, J., & Norman, D. (1994). Representations in distributed cognitive tasks. *Cog. Sci., 18*(1), 87-122.

Zhang, J. (1997). The nature of external representations in problem solving. *Cog. Sci., 21*, 179-217.